\documentclass{aa}

\begin{document} 
\input psfig.tex 
\title{The buildup of stellar mass and the 
3.6 $\mu$m luminosity function in clusters from $z=1.25$ to $z=0.2$.}
\author{S. Andreon}
\institute{INAF-Osservatorio Astronomico di Brera, Milano, Italy
e-mail: andreon@brera.mi.astro.it} 
\date{Accepted ... Received ...}

\titlerunning{The buildup of stellar mass function}

\abstract{

We have measured the 3.6 $\mu$m luminosity evolution of about 1000 galaxies in
32 clusters at $0.2<z<1.25$, without any a priori assumption about luminosity
evolution, i.e. in a logically rigorous way.  We find that the luminosity of
our galaxies evolves as an old and passively evolving population formed at high
redshift without any need for additional redshift-dependent evolution. Models
with a prolonged stellar mass growth are rejected by the data with high
confidence.  The data also reject  models in which the age of the stars is the
same at all redshifts.  Similarly, the characteristic stellar mass evolves,
in the last two thirds of the universe age, as expected for a stellar
population formed at high redshift. Together with the old age of stellar
populations derived from fundamental plane studies, our data seems to suggest
that early-type cluster galaxies have been completely assembled at high
redshift, and not only that their stars are old. The quality of the data allows
us to derive the LF and mass evolution  homogeneously over the whole redshift
range, using a single estimator. The Schechter function describes the galaxy
luminosity function well. The characteristic luminosity at $z=0.5$ is is found
to be 16.30 mag, with an uncertainty of 10 per cent.

{}}  

\maketitle

\section{Introduction}

The luminosity function (LF) is the basic statistic  used to understand
galaxy properties, giving  the relative frequency of galaxies of a given
luminosity in a given volume.  Most additional parameters determined for 
samples of galaxies having more than a single value of luminosity are
usually averages weighted by the LF (in addition to underlying selection
effects).  Furthermore,  by comparing the LF at different redshifts or
environments it is possible to infer how galaxy luminosity evolves. 

Since starlight at 3.6 $\mu$m very nearly follows the Rayleigh-Jeans limit
of blackbody emission for $T > 2000$ K, the colors of both early- and
late-type stars are similar.  There is virtually no dust extinction at this
wavelength either, since any standard extinction law predicts only a few
percent of the extinction of optical wavelengths. The 3.6 $\mu$m
light therefore traces the stellar mass distribution free of dust
obscuration effects (Pahre et al. 2004). Thus, a useful approach to
understanding how galaxies form is to track their growing stellar mass,
measured through the evolution of the 3.6 $\mu$m LF.

Several previous studies addressed the luminosity evolution of galaxies in
clusters in near-infrared bands, notably de Propris et al. (1999). They
found that the $K$-band LF of 38 clusters up to $z=0.92$ is consistent with
the behavior of a simple, passive luminosity evolution model in which
galaxies form all their stars at high redshift and thereafter passively
evolve. However, and perhaps because a standard cosmology was not in
place at the time of that work, the authors do not address which of
the other possible evolutionary scenarios are rejected by the data and at
what confidence. Kodama \& Bower (2003), Kodama et al. (2004) found 
results compatible with little evolution in stellar
mass, but with large uncertainties.

The de Propris et al. (1999) results were anticipated (and also
later confirmed) by the analysis of sub-samples of cluster galaxies:
early-type or red galaxies have properties (mainly colour and location
in the fundamental plane) consistent with a passive evolution model.
However, red or early-type galaxies are biased subsamples of the
whole galaxy population and are affected by the progenitor
bias  (van Dokkum \& Franx 2001). Furthermore, clusters of galaxies
are also know to host star forming galaxies (e.g.
Butcher \& Oemler 1985). The  fraction of
blue galaxies growing with redshift 
(see e.g. Butcher \& Oemler 1985; Rakos \& Shombert
1995; but see Andreon, Lobo \& Iovino 2004 for a different opinion)
makes results based on the red/early-type population less 
representative of the whole population as redshift increases.
LF studies do not suffer from
the progenitor bias, and they directly approach the more fundamental
problem of studying the evolution of the whole sample of galaxies.
Therefore, LF studies are preferable to measure the ensemble mass
evolution. 
Furthermore, stellar populations may be old,
as pointed out by several colour or fundamental plane studies, but at the
same time galaxies may not have been completely assembled: since $z=1$
and the present-day, galaxies might have grown in mass through mergers. This key
issue can be tested by comparing the mass function of galaxies 
at different redshifts.

The paper is organized as follow: Sec 2 presents the data;  Sec 3
describes the sample of studied clusters. Sec 4 summarizes the
method used to derive the LF. The LF and the mass evolution determination
are computed in Sec 5. Sec 6 and 7 discuss and summarize the
results.

Throughout this paper we assume $\Omega_M=0.3$, $\Omega_\Lambda=0.7$ 
and $H_0=70$ km s$^{-1}$ Mpc$^{-1}$.

\section{Data \& data reduction}

\subsection{IR data}

IR data were obtained with the IRAC (Fazio et al. 2004) on the Spitzer
Space Telescope (Werner et al. 2004). A 9 deg$^2$ SWIRE (Lonsdale et
al. 2003) field was imaged. The exposure time is 4 $\times$ 30s. 

The standard pipeline pBCD (Post Basic Calibrated Data, ver. 10.5.0)
products delivered by the Spitzer Science Center (SSC) were used in
this paper. These data include flat-field corrections, dark
subtraction, linearity and flux calibrations. Additional steps
included pointing refinement, distortion correction and mosaicking.
Cosmic rays were rejected during mosaicking by sigma-clipping. 
pBCD
products do not merge together observation taken in  different
Astronomical Observations Requests (AORs). AORs are therefore mosaicked
together using SWARP (Bertin, unpublished), making use of the weight maps.
Sources are detected using SExtractor (Bertin \& Arnout 1996), making
use of weight maps.

Star/galaxy separation is performed by using the
stellarity index provided by SExtractor, which, being 
the posterior probability based on a neural
network, outperforms cuts (linear discriminators) in object
parameter space (Andreon et al. 2000), as expected (e.g. Bishop 1995).
We conservatively keep a high posterior threshold ($class_{star}=0.95$),
rejecting ``sure star" only ($class_{star}>0.95$)
in order not to reject galaxies (by unduly putting them in the star class),
leaving some residual stellar contamination in
the sample. This contamination is later dealt with statistically.

We checked the star/galaxy classification using a 0.3 deg$^2$ region
deeply observed at Cerro Tololo
Inter--American Observatory (CTIO) by taking images 
under sub-arcsecond seeing conditions (see
sect 2.2). We compared the classification derived  from 3.6 micron 
images with
those of one of our CTIO data observed with sub-arcsec resolution.
Only less than about 2 per cent of the objects classified as stars (using
$class_{star}>0.95$) using IRAC images are actually resolved at the
CTIO resolution,  and are therefore mis-classified. This 2 per cent of stars 
corresponds to 
0.1 per cent of all the sources detected at 3.6 $\mu$m in the same field (and
brighter than 18 mag), i.e. a negligible minority overall. Therefore,
our posterior probability threshold ($class_{star}>0.95$) does not
reject galaxies by unduly putting them in the star class at an
appreciable level. The residual stellar contamination from
our choice of the posterior threshold is subtracted 
statistically, as in the optical (e.g. Andreon \& Cuillandre 2002) or
the near--infrared (e.g. Andreon 2001). 

Images are calibrated in the Vega system, using the IRAC zero points
provided by the SSC (and, in particular by G.
Wilson\footnote{http://spider.ipac.caltech.edu/staff/gillian/cal.html}.

From the inspection of the galaxy count distribution,  the completeness
magnitude  at $[3.6]$ is $\sim 18$ mag. Objects brighter than 12.5 mag are often
saturated. Therefore, from now on, only the range $12.5< [3.6] < 18$ mag is
considered. In sec. 5.3 we check how 
results are affected by a potential incompleteness at the faint end.

The average density of sources in a circle of the 
point spread area is around 0.004. Therefore, it is very unlikely that
crowding is an issue in average density regions, and also
in 10 to 100 times overdense regions, such as cluster cores.
Only one of our clusters
required an accurate setting of the SExtractor deblending parameters
to split the few blended sources.

\subsection{Optical imaging data}

In this paper we used CTIO wide-field imaging to
control the quality of Spitzer star galaxy classification, as mentioned above.
We adopt here part of the same imaging data used in 
Andreon et al. (2004a).   In brief, optical $R$-- and $z'$--band
($\lambda_c\sim9000${\AA}) images were obtained at the CTIO
 4m Blanco telescope during  
August 2000 with the Mosaic II camera. Mosaic II is a 8k$\times$8k 
camera with a $36\times 36$ arcminute field of view. 
Typical exposure times were 1200
seconds in $R$ and $2 \times 750$ seconds in $z'$. Seeing in the final
images was between 0.9 and 1.0 arcsecond Full--Width at Half--Maximum
(FWHM). Data have been reduced in the standard
way (see Andreon et al. 2004a for details). 
Typical completeness magnitudes are $R=24.5$, $z'=23$ mag
($5\sigma$) in a 3 arcsec aperture.

\section{The cluster sample}

\begin{figure}
\psfig{figure=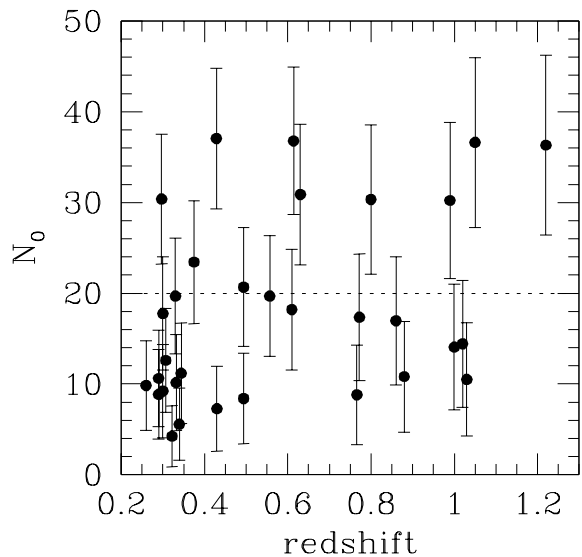,width=8truecm}
\caption[h]{
Richness for our clusters. The dotted line marks the typical $N_0$
of $R=0$ or $1$ cluster in the present day Universe, from Bahcall (1977).
} 
\end{figure}

The cluster sample studied in this paper consists of 32 colour--selected
clusters, all spectroscopically confirmed. The clusters were detected as
spatially localized galaxy overdensities of similar colour,  as described in
Andreon et al. (2003; 2004a,b). The detection method used takes advantage of
the observation that most galaxies in clusters share similar colours, while
background galaxies have a variety of colours, both because they are spread
over a larger redshift range and because the field population is more variable
in colour than the cluster one, even at a fixed redshift.  All clusters but one
were detected using $R-z'$ colour; one cluster was detected using the $z'-K$
colour because of its larger redshift (sec 3.2 of Andreon et al. 2005 for
details about the latter cluster).   After colour detection, the cluster nature
of the studied clusters was confirmed with  spectroscopic observations. The
clusters are individually presented and studied in Valtchanov et al. (2004),
Pierre et al. (2005), Willis et al. (2005),  Andreon et al. (2004a, 2005a,b).

Altough studied clusters
are colour--selected, 29 out 32 of the clusters are also 
x-ray detected, leaving only 3 clusters (at $z=0.49, 0.61$ and at $z=1.02$) 
with too faint x-ray emission to be detected with 10 to 20 ks XMM images.
The detected x-ray emission guarantees that the studied
clusters have deep potential wells and independently confirms
the cluster nature of the studied objects.

The studied cluster sample is not a volume complete sample, 
nevertheless it densely
samples the explored Universe volume, up to $z\sim1$.
Assuming the local (Ebeling et
al. 1997) $L_X$ luminosity function, we found that the expected number of
clusters with x-ray luminosity $L_X> 10^{43}$ erg/s up to $z\approx 1$ and
with more than 50 counts on 10 ks XMM images is about 15 per deg$^2$.
All 32 studied
clusters are in a contiguous 2.8 deg$^2$ area of the sky, and
therefore the cluster number density is
about 11 clusters per deg$^2$. The high number density of clusters 
should make the studied sample
somewhat representative of typical clusters, 
making any bias of the studied sample small.

In order to characterize the cluster richness,
for each cluster we compute the number of galaxies inside a radius of
357 kpc (corresponding to 500 kpc in the old cosmology and
in the nearby universe studied by Bahcall 1977) brighter than
$m^*+1.7$ (corresponding to Bahcall $\sim m_3+2$ for an $\alpha=-1$
Schechter function). The background contribution has been removed
(marginalized) using Bayesian techniques described in Appendix
B of Andreon et al. (2005). Here we quote 
the mean of the posterior and its rms as a measure of
the cluster richness and its error, respectively, for a uniform 
prior (but results are similar using a Jeffreys prior).
Clusters of Abell (1958) richness 0 
or 1 in the present--day Universe have $N_0$ of about 
15 to 20 (Bahcall 1977). 
Figure 1 shows that two thirds of
our clusters have $N_0<20$ galaxies. Most of the clusters
with $N_0>20$ galaxies have large error bars, which
make them consistent with $N_0=20$ galaxies at
two sigma. Therefore, most of our clusters are at the bottom
of the Abell richness scale and are not rich systems.

The 32 clusters are distributed in the redshift ranges in the following
way (see also Figure 2): 13, 8, 5, 6 clusters are in the range: $0.25< z <0.40$,  
$0.40< z < 0.65$, $0.75< z <0.90 $ and $0.99< z <1.25$, respectively.

\begin{figure}
\psfig{figure=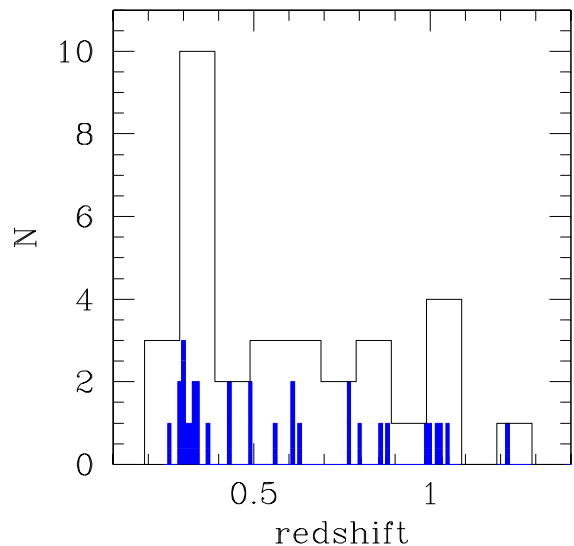,width=8truecm}
\caption[h]{
Redshift distribution of the studied clusters. Solid/open histograms mark
0.01/0.1 bin width in redshift.}
\end{figure}

\section{The method: LF determination}

We do not attempt here, as in some previous approaches, to infer the
luminosity function from an optically selected sample, because the
latter option assumes the absence of very dust obscured
objects:
if galaxies are dust obscured
enough to go undetected in the optical, their contribution to the IR LF
is deemed to be zero, when instead it may be relevant. Some previous authors
were forced to assume the absence of very dust obscured
objects, because they need spectroscopic
redshifts, mostly acquired in the optical window,  or because
they need multi--colour
optical photometry to determine a photometric redshift. Here,
instead, we use the approach usually adopted for clusters,
where knowledge of the individual galaxy redshift is not used.

\begin{figure}
\psfig{figure=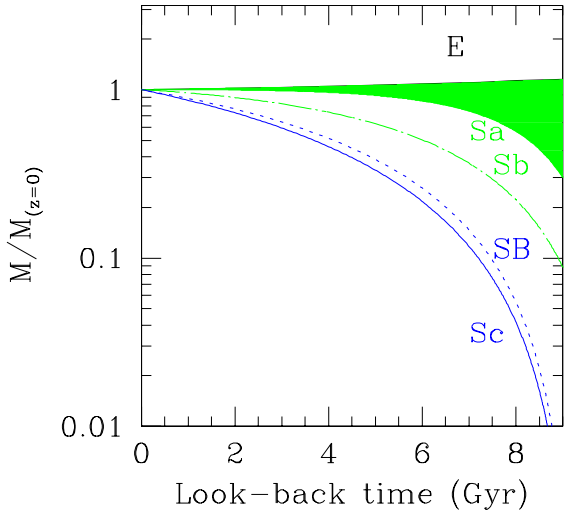,width=8truecm}
\caption[h]{Stellar mass evolution for several stellar mass growth models 
(curves) and as derived from our data (shaded area).  
The region allowed by the data is part of the shaded green area
derived in sec 5.2.}
\end{figure}

In order to compute the LF we adopt two different methods.

-- For display purposes only, the LF is
computed as the difference between the (binned)
counts in the cluster and control
field direction, as usual (e.g. Zwicky 1957, Oemler 1973). 
Error bars are computed as the square root of the
variance of the minuend, because the contribution due to the
uncertainty on the true value of background counts is negligible. 
A negative number of cluster
galaxies may occur in the presence of a low
cluster signal and Poissonian fluctuations,
which leads to the unphysical result of negative
numbers of cluster galaxies when data are binned.

-- Second, we fit the unbinned galaxy counts, without any use
of binned data or errors computed in the previous approach. We are
faced with the classical statistical problem of
determining two extended  (integral$>1$) density probability function, one
carrying the signal (the LF of cluster members) and the other being due to a
background (background galaxy counts, BKG) from the observations of many
individual events (the galaxies luminosities), without knowledge of which
event is the signal (which galaxy is a member) and which one is background.  
Here, we follow the rigourous method set forth in Andreon, Punzi \& Grado 
(2004, APG hereafter), which is an
extension of the Sandage, Tammann \& Yahil method (1979, STY) to the
case where a background is present, and that adopts the  extended
likelihood instead of the conditional likelihood used by STY. The
method does not remove the background from the data, but adds a  component
(the background) to the model. The method also provide the
normalization (at the difference of STY), needed to rigorously combine
the LF of the individual clusters, properly accounting for the
uncertainty in the LF normalizations. 68 per cent confidence
intervals are derived using the Likelihood Ratio theorem (Wilks 1938,
1963) made known to the astronomical community by Avni (1976) and Press
et al. (1986), among others. 
The method comes in two forms: in
sec 5.1 we neglect astronomical and statistical subtleties, proceeding
in the analysis as most previous published papers, whereas in sec
5.2 we take a fully rigorous approach, and we describe its advantages
with respect to the simpler application. 

\begin{figure*}
\centerline{%
\psfig{figure=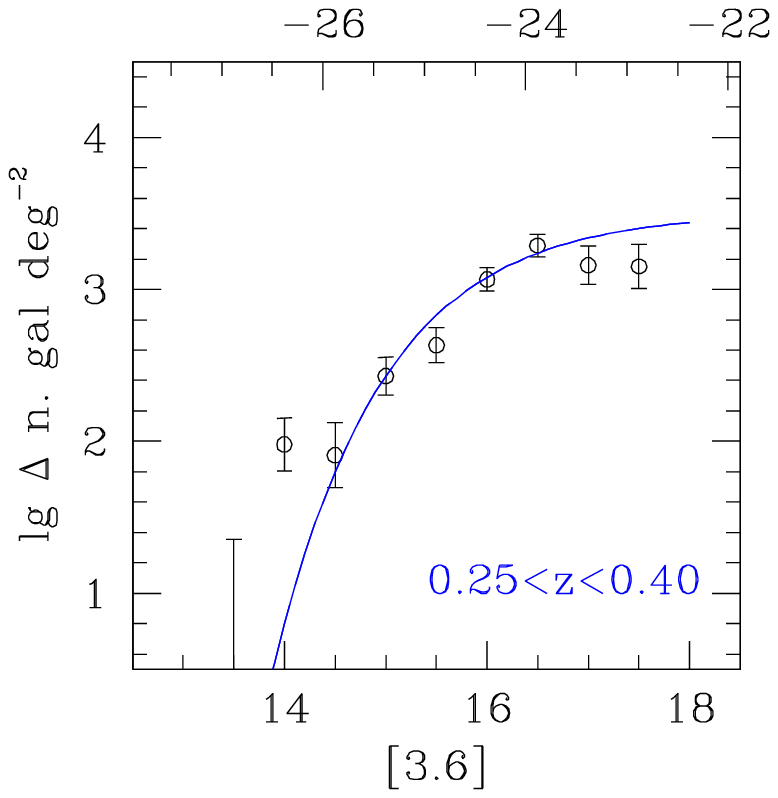,width=6truecm}
\psfig{figure=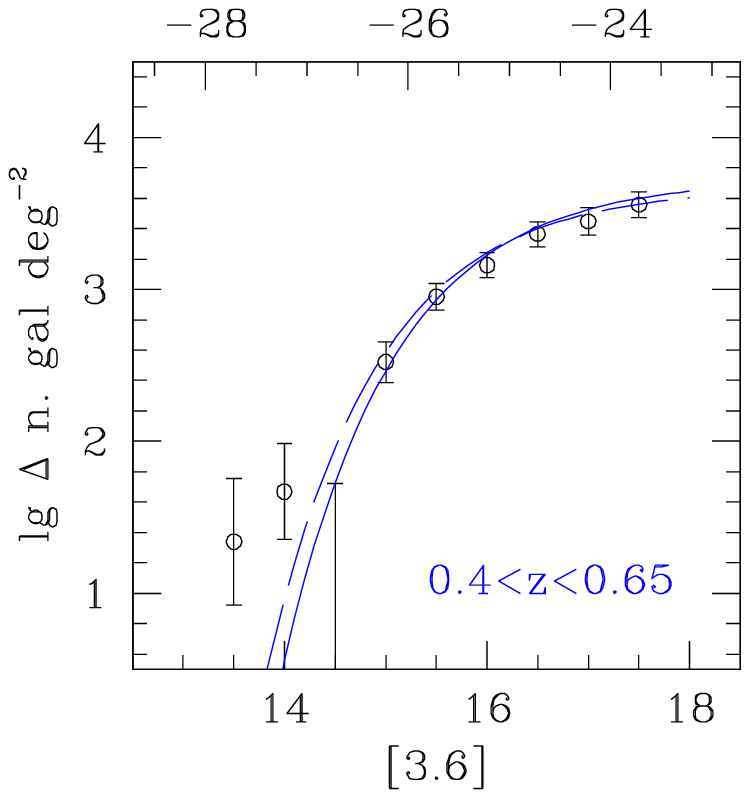,width=6truecm}}
\centerline{%
\psfig{figure=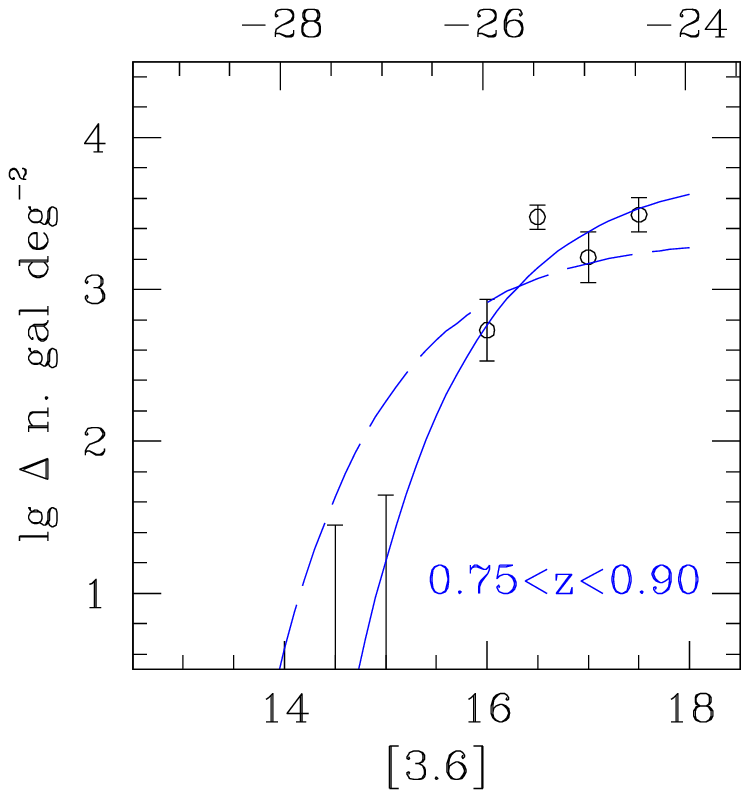,width=6truecm}
\psfig{figure=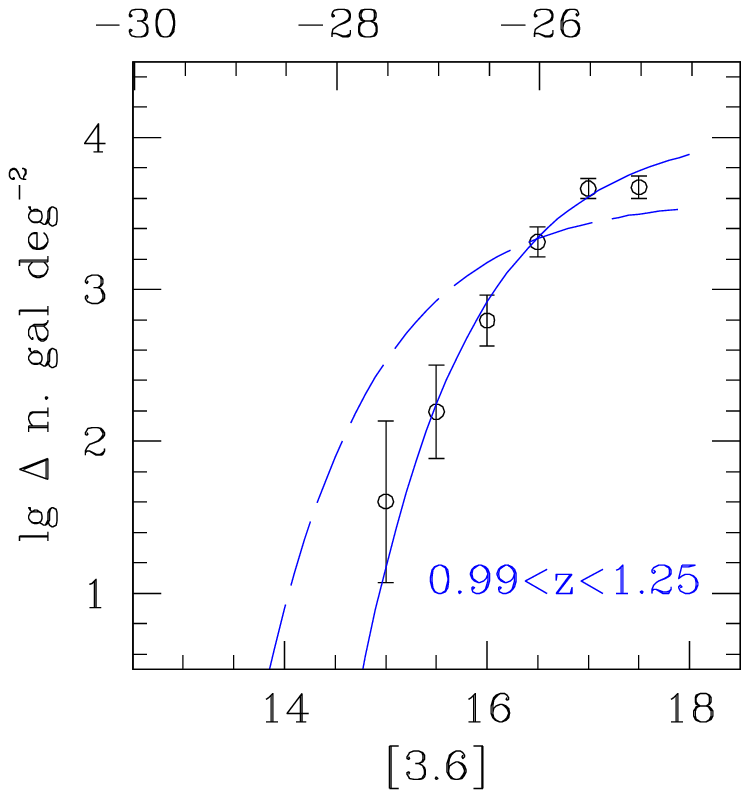,width=6truecm}}
\caption[h]{
Composite LF in the $[3.6]$ band as a function of
apparent (lower abscissa) and absolute (upper abscissa) magnitudes.  Data points and error bars are
computed as usual (e.g. Oemler 1973, see sec 4). The solid and long dashed
curves are fit to unbinned counts, but neglecting astronomical and statistical
subtleties, as described in sec 5.1. The solid curve refers to a fit with
$m^*$ free, whereas the dotted curve is a fit with $m^*$ held fixed to the
value observed at $0.25< z <0.40$. All these LF determinations are
superseded by
the LF determined in sec 5.2, which is a rigorous fit over all cluster and
field data.}
\end{figure*}

\subsection{Background and cluster areas}

As the background field we considered a central 4 deg$^{2}$ region
for simplicity, and we fit the background counts
with an arbitrary function.  In this region, there are
about 106000 objects.  The {\it average} background is therefore
very well determined, because it is measured over a large area with respect to
the cluster area. Its determination is so good that galaxy counts in
the cluster direction (an area about 750 times smaller) does not
constrain the background counts at all.  Therefore, it is 
justified here to keep the background parameters at the best fitting values
observed in the control field (but see APG for why this is unjustified in
general). 

For all but 2 clusters, we measured the LF within a circle of 5 
arcmin aperture. For the remaining two clusters 
an aperture of 3 arcmin is taken because of incomplete data
coverage or because of a bright nearby star. This 
aperture is similar to the one
used in the optical LF determination for
several clusters in common with
Andreon et al. (2004a), and it has
been chosen as a compromise between sampling the whole cluster and
not including a too large contribution from background galaxies.

\subsection{Evolutionary models}

In order to estimate the expected apparent magnitude, absolute
magnitude and mass to light ratio for different
galaxy models having various
growth histories we used GRASIL (Silva et al.
1998; Panuzzo et al. 2005), which is a code to compute the spectral
evolution of stellar systems taking into account the effects of dust,
which absorbs and scatters optical and UV photons and emits in the
IR-submm region. We adopt standard 
elliptical (E), Sa, Sb, Sc and Arp 220 (SB) models with default
parameters (Silva et al. 1998): 
a Salpeter initial mass function is used,
with lower/upper limit fixed to 0.15/120 $M_\odot$.
We assume that no stars are formed from $z=\infty$ to
$z=5,2,2,1.5,1.5$ for E, Sa, Sb, Sc and SB models, 
respectively.
The models fully account for evolving metallicity and dust content
with dust mixed to stars (see Silva et al. 1998 for details).
Figure 3 shows model mass growth histories
appropriate for an object having currently
broad band spectrophotometry typical of E, Sa, Sb,
Sc and star burst (SB) galaxies.  The E model of stellar mass growth history is characterized
by the absence of recent stellar mass growth, while later 
types display recent episodes of stellar mass growth. 
These stellar mass growth histories are not intended to
represent the stellar mass growth history of an individual object,
but only of the average class, which is why
these curves are smooth, while the mass growth of individual object
is more erratic.

\section{Results}

\subsection{A simplistic approach}

We start with a simple analysis of the data, without 
statistical and astronomical subtleties. We largely
follow the usual astronomical method of LF computation {\it in the
field} and also in the cluster when many clusters are available. First, we bin
data in redshift bins. In this step we ignore that the redshift bin is
of non--vanishing width (i.e. we overlook the required convolution of
the model by the appropriate redshift kernel) and that sources likely
brighten in their rest--frame due to the younger age of stars going
from the near to the far side of the redshift bin. We also overlook
some statistical subtleties (each cluster has a Schechter function, not
just the composite), and logical coherence (we are attempting to
measure the evolution assuming its absence,
as most previous works did, and as criticized by Andreon 2004).  A
fully rigorous analysis is presented in section 5.2.

There are 260, 250, 100 and 240 cluster (i.e. background subtracted)
members inside the composite clusters at $0.25< z <0.40$,  
$0.40< z < 0.65$, $0.75< z <0.90 $ and $0.99< z <1.25$, respectively.

As a model for the cluster LF we adopt a Schechter (1976) function:

\begin{equation}
\phi_i(m)=\phi^*_i 10^{0.4(\alpha+1)(m^*-m)}e^{-10^{0.4(m^*-m)}}
\end{equation}

where $m^*$, $\alpha$ and $\phi^*_i$ are the characteristic magnitude,
slope and normalization, respectively. The index $i$ refers to the
cluster $i^{th}$.
In this section, we fixed the Schechter slope $\alpha$ to $-1$ because
it is largely undetermined.

\begin{figure}
\psfig{figure=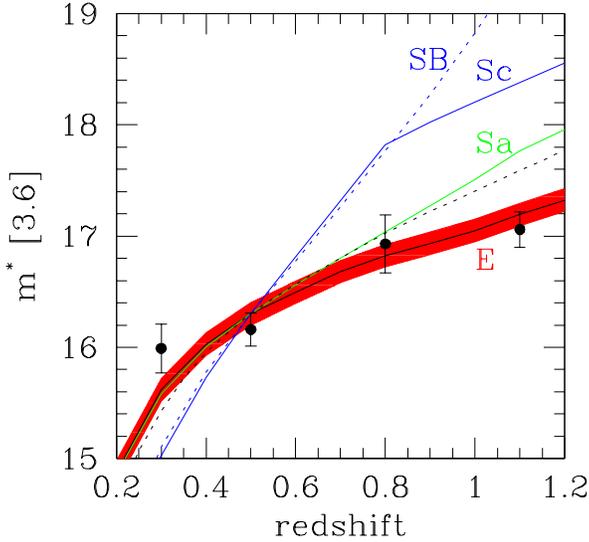,width=8truecm}
\caption[h]{The different lines shows the
apparent magnitude of a galaxy having $[3.6]=16.3$ mag
at $z=0.5$, depending on stellar mass growth histories, as labelled,
as well as for an unphysical model where galaxies have the same 
old age at all redshifts  (dotted black line). 
Data points are $m^*$ values derived in sec 5.1. The red region marks
the {\it observed} 68 per cent confidence bound of $m^*$ derived in sec 5.2. 
}
\end{figure} 

Figure 4 shows the LF in each redshift bin. 
Data points and error bars are computed as usual 
(e.g. Oemler 1973). When the difference between the cluster and control field
counts is negative the result cannot be  plotted, because the logarithm
function requires a positive argument.  The
solid curve is the LF of the composite dataset, obtained by fitting the
unbinned counts in the composite cluster and control field directions.
We do not fit the displayed data points shown in
the figure, and the LF fit (parameter or error determination) makes no
use of these data points and errors, and they are shown for display
purposes only: we fit, as mentioned, unbinned counts in the cluster
and control field directions.

The `data' points nicely follow the Schecther function. As redshift
increases, the LF moves to the right, i.e. $m^*$ becomes fainter in the
observer frame,  by about 1.0 mag between $0.25< z <0.40$ and $0.99< z
<1.25$. In all redshift panels also plotted, as a long-dashed curve, is a fit
with $m^*$ held fixed to the value observed in the lowest redshift bin.
High-redshift data reject a model with $m^*$ held fixed to the value observed
in the lowest redshift bin.

Figure 5 shows the expected apparent magnitude of a galaxy having
$[3.6]=16.3$ mag at $z=0.5$, for some stellar
mass growth histories. It shows that a minimum of 1.5 mag of fading is expected in
apparent magnitudes going from $z=0.3$ to $z=1.1$, whereas only
1.0 mag of fading is observed. However, it is not obvious from this
figure whether the data reject the various stellar mass growth histories 
and at what confidence. We
will not pursue the statistical computation using the approximate
method just described, but using the rigorous method detailed in the
next section.

\subsection{Adding rigour}

Deriving luminosity evolution after having assumed that it is equal to
zero in each studied redshift bin (i.e. the approach of the previous
section) makes use of a circular argument:
the computation of the LF assumes a model for galaxy
evolution, that unfortunately is precisely what the LF is used to
measure. 
This occurs even for the traditional manner in which the LF is computed:
having observations at different redshifts (look-back times) and
desiring to measure how galaxies evolve, we should count galaxies
having a given absolute mag at, say, $z=1$ with galaxies having  the {\it same}
absolute mag at, say, $z=0$ (and in such a case we assume that galaxy
luminosity does not change with look-back time), or with galaxies
having a {\it different} luminosity (and in such a case we assume a given
evolution). And, if we know how the luminosity evolves (because this is
needed to make the computation), there is no need to perform the experiment 
(why measure the LF?). 

Beside the logical inconsistency, the assumption of
a level of evolution underestimates error bars by a significant factor, as
measured in an actual case by Andreon (2004). 
We stress that a rigorous method is required if one wants to be sure of the
correctness of the result, and, when the number of members in a given mag
bin, estimated from the difference of galaxy counts in the
cluster and control field directions, leads to unphysical (negative)
values. One should, instead,  
solve for background counts, for the LF
and its evolution at the same time, {\it without binning
the sample in redshift bins} and without bin the data in magnitude bins. 
This can be achieved adopting
a more complex model in which additional parameters account for the
evolution, following the path in previous works 
(Lin et al. 1999, Blanton et al. 2003, Andreon 2004) and 
reinforced in APG. 

Algeabrically, $m^*$ in Eqn. 1 is replaced by:

\begin{equation}
m^*_z=m^*_{z=0.5}+\Delta m_{model}-Q \ (z-0.5)
\end{equation}

where $\Delta m_{model} = m_{model} - m_{model, z=0.5}$.

Thus, we make the $m^*$ model fainter by an amount given by the
prediction appropriate for the adopted stellar mass growth history 
and cosmology and we
allow a supplementary linear evolution (i.e. proportional to redshift),
normalizing everything to $z=0.5$ to make corrections equal to zero
at roughly the median redshift of our survey. 

Our $Q$ has a different meaning from $Q$ in
Lin et al. (1999), Blanton et al. (2003) and Andreon (2004): here
$Q=0$ means that galaxies form stars
according to the model, whereas in the
former works $Q=0$ means that the luminosity stays constant over time,
which is, of course, unphysical. 

The sample consists of about 5500 galaxies, of which about 950 galaxies
are in clusters, and the remaining are interlopers. 

\begin{figure*}
\centerline{%
\psfig{figure=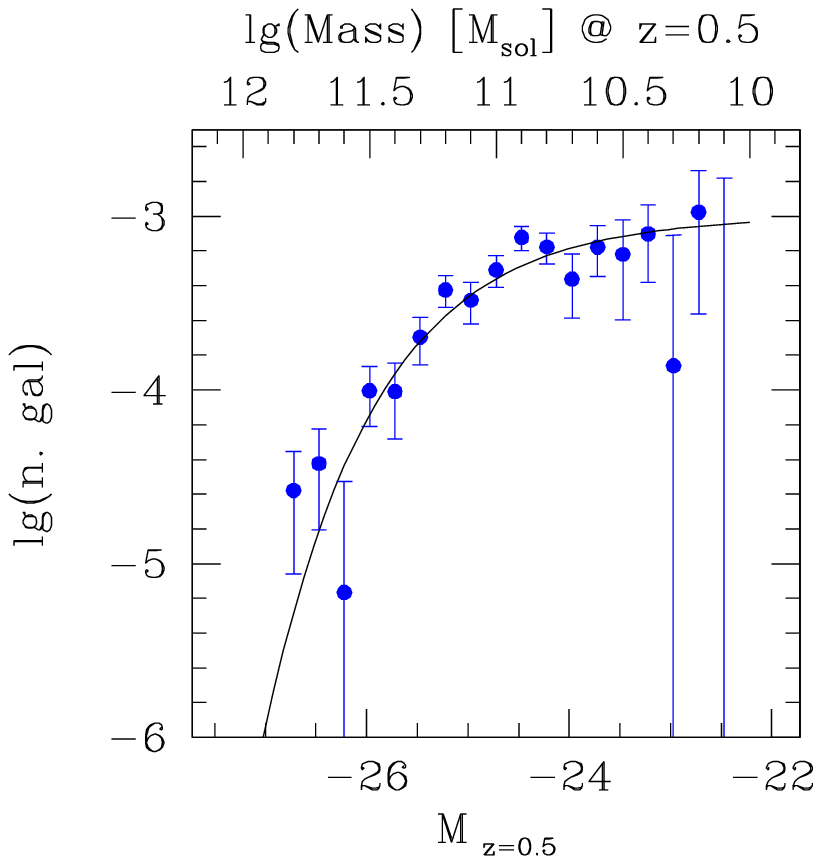,width=8truecm}
\psfig{figure=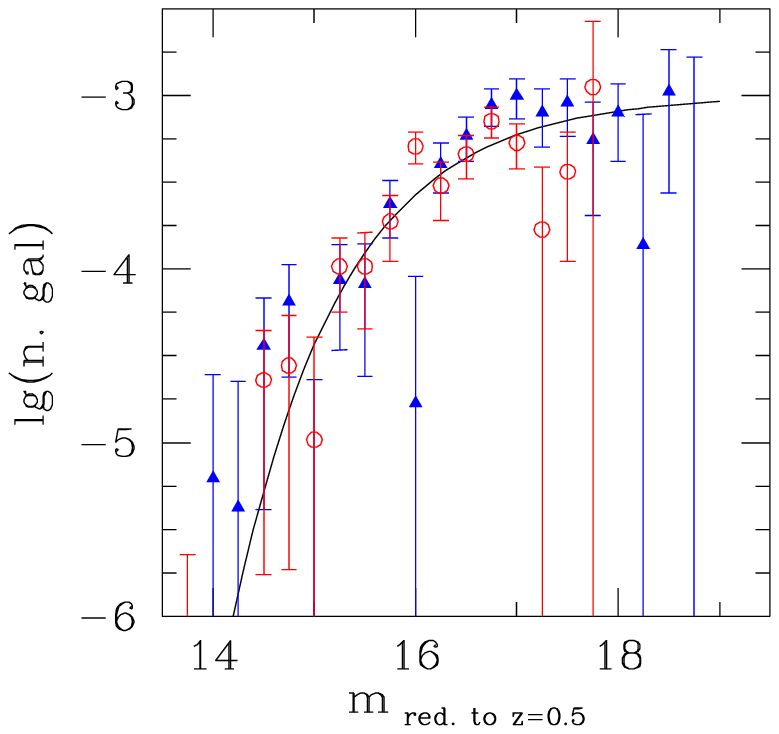,width=8truecm}}
\caption[h]{
Best fit LF at $z=0.5$. {\it Left:} LF of the whole sample.
{\it Right:} LF of the lower and upper redshift halves
(blue closed triangles and red open circles, respectively).
The match of the two halves occurs only if the assumed
evolution is the correct one.
}
\end{figure*}

To compute the LF, we start with model
selection. We use two statistical tests:  the
likelihood ratio test (LRT) will inform us of how frequently one incorrectly
rejects the null hypothesis, under the hypothesis that the null is true.
The Bayesian Information Criteria (BIC, Swartz 1978; Lindle 2004
provides a useful astronomical introduction) informs us about the relative
evidence of two models. LRT cannot be used when regularity conditions
required for its application do not hold (for example, compared models
should be hierarchically nested, i.e. one model should be a particular case 
of a more general model).

We compare the E model to the other models, all with fixed
$Q=0$, irrespective of the $m^*$ and $\alpha$ values. BIC provides very
strong support for the E stellar mass growth history ($\Delta BIC\ga5$ for Sa,
and larger values for the other models). 
Our  statistical analysis offers the advantage of
avoiding an unnecessary assumption about the value of the best fit
parameters in order to identify the most likely evolutionary model: 
it is able to 
infer an evolutionary model without any assumption about the $\alpha$
value, while such an assumption was done in sec 5.1 and in previous
works, because these studies were forced to fix the
$\alpha$ parameter to derive the evolution of $m^*$. Not fixing
the $\alpha$ value, our approach is not affected by
the known correlation between $m^*$ and $\alpha$ that
plagued previous approaches. 

Figure 3 summarizes the above model comparison:
the acceptable area (i.e. the constraint put on models by our data) is a part
of the shaded (green) area plotted in Figure 3, ie. the region between the E
track (a good description) and the Sa track (a bad one). 
The major difference between the Sa track
(rejected by the data) and E track begins at look--back times greater than
7 Gyr, i.e. at $z>0.85$, and becomes large at look--back times greater than
7.5 Gyr (i.e. $z\ga1$). We can discriminate 
between the two models because we have 8 clusters at $z>0.85$ and 
5 clusters at $z>0.99$, where models differ the most (Fig 3).

We now verify whether
the model is a) too complex (too many
unconstrain parameters) for the
data in hand, and b) if the data requires that the E model should be
updated with a better one. We compare the E model having
fixed $Q=0$ and $\alpha=-1$ to more general E models with $Q$ or $\alpha$
free.  
BIC and LRT both inform us that the simplest model is favored. Given
the data in hand, the E model does not need to be refined by the addition of a
linear (with redshift, i.e. a $Q$) term. Furthermore, the $\alpha$
parameter is largely unconstrained, quantifying what is 
qualitatively apparent in Fig. 4.

BIC strongly rejects the unphysical universe (the track of an E
with the present-day age at all redshifts).

The final best fit model is therefore the one of an old and passively
evolving population formed at high redshift, without any additional
recent stellar mass growth (i.e. the E model and $Q=0$). For
$\alpha=-1$ (the best formal fit is $\alpha=-1.05$, but with large
error bars),  we found $m^*_{z=0.5}=16.30\pm0.10$ mag,
i.e. $M^*=-24.8,-24.9,-25.1,-25.3$ mag at $z=0.3,0.5,0.8,1.1$.

Figure 6 shows the rigorously derived best fit LF (the curve) at $z=0.5$ and  
the data points. Having measured the
luminosity evolution, we can now safely
assume it in order to combine counts at different redshifts (look--back
times). We normalize each individual LF by the model
$\phi^*_i$  value, and we then weight each cluster by its $\phi^*_i$
value, after having evolved magnitudes from the cluster redshift to our
reference redshift, $z=0.5$.  In doing this convolution we keep only bins
entirely included in the studied magnitude range, for simplicity.  As in
literature approaches, errors on the data points do not include
data combining errors (whereas they are accounted for in our
rigorous derivation, see APG). There is a good agreement between the
curve and the data points, meaning that the Schechter function is a
good description of the LF over the observed magnitude range, and that
the selected stellar mass growth history provides a good description of the
observed evolution. The latter point is displayed in the right panel
of Figure 6. It shows the LF by splitting the sample in two redshift
halves, at median redshift. The two LFs share a common $m^*_{z=0.5}$,
and this occurs only if we selected the right stellar mass growth history
model: if the adopted model underestimates the luminosity evolution in
one of the halves, two different (horizontally shifted) LFs would be
observed (as we discuss later). 
This ``result" is a visual check of 
the model selection already discussed. 

Figure 7 qualitatively shows why we have ruled out the evolution of a stellar
population whose stellar mass growth history is appropriate for an Sa (or
later types): it displays the LF of clusters at $z<0.85$ and
$z>0.85$, separately, evolved to $z=0.5$ assuming the stellar mass growth
history of an Sa population. The $m^*_{z=0.5}$ of the two samples do not match,
as emphasized by the bottom panel that shows  the cumulative LF:  the
high-redshift LF is left-shifted (too bright) with respect to the
low-redshift LF, contrary to the adopted hypothesis that
evolution is well described by the adopted stellar mass growth history. 
The mass evolution allowed by the data is much
quieter than that of an Sa galaxy. The model selection 
performed confirms this result rigorously, quantifying 
its statistical
significance.

\begin{figure}
\psfig{figure=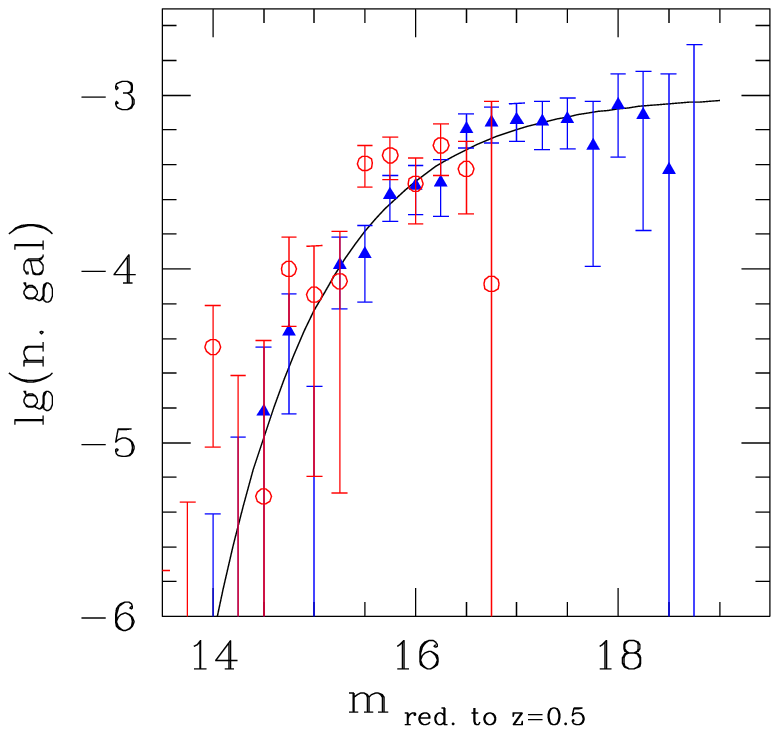,width=8truecm}
\psfig{figure=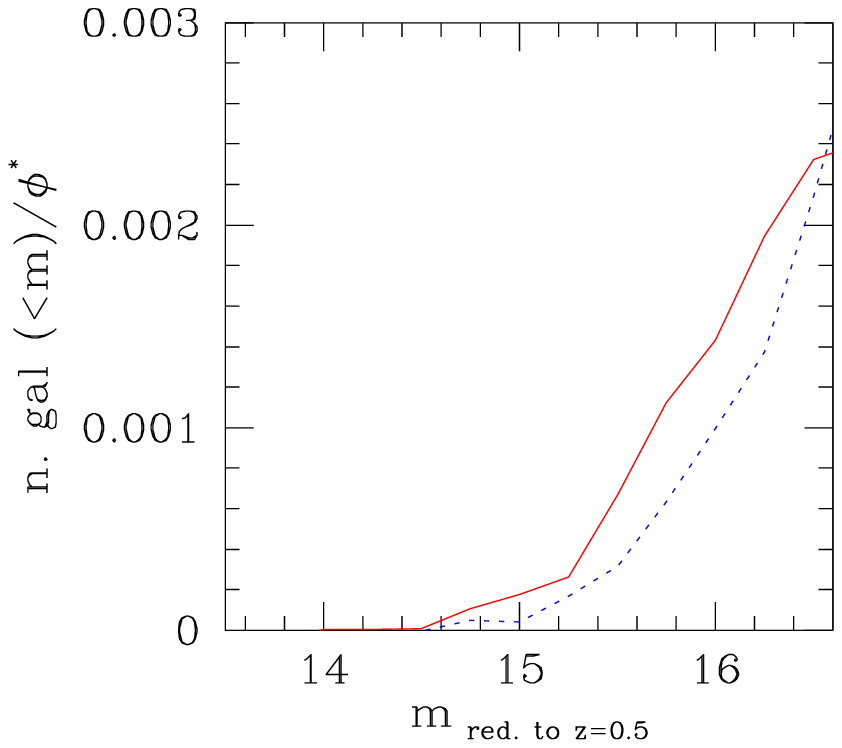,width=8truecm}
\caption[h]{
{\it Top panel:}
LFs, evolved to $z=0.5$ assuming an Sa stellar mass growth history, of the lower and
upper redshifts ranges (marked by blue closed triangles and red open circles for
$z<0.85$ and $z>0.85$, respectively). The mismatch 
between the two LFs (circles are
left-shifted) implies that
the assumed stellar mass growth history is rejected by the data. 
{\it Bottom panel}: blow-up of the cumulated LF, better
showing the LF mismatch.
}
\end{figure}

Having identified the evolutionary model, we can convert
absolute magnitudes into a stellar mass scale, using the $\mathcal{M}/L$
ratio of the model. The mass scale is shown as the upper abscissa
in Fig 6. We found
$\mathcal{M^*}=1.1,1.2,1.2,1.2 \ 10^{11} M_{\odot}$
at $z=0.3,0.5,0.8,1.1$, respectively, were 
$\mathcal{M^*}=M^* (\mathcal{M}/L)_{model}$ (in the appropriate units). 
The statistical accuracy is 10 per cent, derived from the 
$m^*$ uncertainty only. The absolute
value of the characteristic mass, $\mathcal{M^*}$, depends on
several key model parameters (e.g. the lower mass limit of the 
initial mass function, see Bell \& de Jong 2001), 
while its evolution does not, as long as model parameters are
redshift-independent.

Thus, we found that the luminosity of our galaxies evolves as 
an old and passively evolving population formed at high redshift. Models
with a prolonged stellar mass growth are rejected by the data with high
confidence. The mass function does not change in the last 8 Gyr, corresponing
to the two thirds
of the current Universe age. The data also reject the need for a redshift-dependent
description of the evolution more accurate than a  passively evolving
population formed at high redshift (i.e. a $Q \ne 0$ is rejected). 
The data also reject models in which the age of the stars is the same at all
redshifts.  The Schechter function well describes the data.

It makes little sense to improve upon these
constraints with our data alone, say to attempt to constrain the 
characteristic time scale of the mass growth
because of a degeneracy: the same luminosity
evolution  may be produced by an older, but longer, episode of star
formation, or a younger but shorter one.

\subsection{Test on incompleteness}

In order to test the effect of a potential incompleteness of the sample at
[3.6] $\sim 18$ mag, we cut the sample at $12.5< [3.6] < 17.5$ mag, and we compute
$m^*$ for the E model and $Q=0$. We
found $m^*=16.54\pm0.11$ mag, in good agreement with the value measured by
considering the larger sample [3.6]$<18$ mag, $16.30\pm0.10$ mag. Application
of BIC still favour the E model over Sa and later type ones, but now at
lower significance ($\Delta BIC \ga 3$) because of the reduced sample size.
Tus, the potential incompleteness of the sample at [3.6] $\sim 18$
mag is too small to affect the results of our analysis.

\section{Discussion}

\subsection{Comparison with previous works}

There are no LFs in the [3.6] band with which we can directly compare,
and therefore, we
compare our [3.6] LF with LFs derived at shorter wavelengths.
Our measure of evolution at 3.6 $\mu$m is a 
luminosity--weighted
measure of evolution. The advantage 
of the chosen band is
that it is far less sensitive to sporadic star formation episodes 
involving a small fraction of the mass than similar 
determinations performed at shorter wavelengths, because
the [3.6] band measures the flux emitted between the $H$ (at $z=1.25$) 
and $K$ (at $z\sim0.1$) band rest-frame. At these wavelengths the
flux-weighted age of a simple stellar population from Bruzual and
Charlot (2003) is about 5 Gyr (e.g. Martin et al. 2005). Instead,
in the B band rest-frame, the
flux-weighted age is  1.5 Gyr, and drops by a factor of 10
at $\lambda \sim 3000$ \AA \ . Therefore, comparison of results 
obtained in different bands requires that we pay attention to the 
considered wavelengths.

Luminosity evolution derived in the [3.6] and $K$ bands can be directly 
compared, because both determinations are sensitive to the same
long-lived stars. Toft et al. (2004) summarized the cumulative efforts
in the literature (largely relying on de Propris et al. 1999) to
determine the LF at $z\ga0.1$. Our data alone match in number and in
redshift distribution this cumulate effort. Toft et al. (2004) and de  
Propris et al. (1999) both find that
the redshift dependence of $m^*$ agrees with that of a passively
evolving population formed at high redshift, as we have found and
as also found by Kodama \& Bower (2003) and Kodama et al. (2004). To
quantitatively compare values derived in different bands, we need to
determine $m^*_{z=0.5}$ for the same slope adopted in the comparison
work ($\alpha=-0.9$, because most of the measurements have been performed with such
a slope), and to convert our $m^*_{z=0.5}$ from [3.6] to the $K$ band.
To perform the latter task we use model $K-[3.6]$ colors provided by 
Grasil. The best fit value converted to the $K$ band is
$m^*_{z=0.5}=16.7\pm0.1$, in good agreement with the value inferred from
Fig. 11 in Toft et al. (2004) $m^*_{z=0.5}=16.6\pm0.5$ mag. Our measure
has a five times better accuracy than the latter, because Toft et al.
(2004) do not fit $m^*$ to their data, but simply force $m^*$ to be
equal to the value observed by de Propris et al. (1998) at $z=0$, 
thus inheriting its accuracy ($\pm0.5$ mag). As mentioned in 
the introduction, our analysis rules out several alternatives
(a non-passive evolution, for example), whereas Propris et al. (1999)
do not address the topic of model selection, perhaps because of the
freedom in cosmological parameters at the time of their 
analysis.

The LF study by Andreon et al. (2004) samples $\approx U$ and $B$
bands and their sample of clusters has a large overlap (13 clusters) 
with the one studied in this paper. They 
found that clusters are composed of two populations, one
that had evolved passively from $z_f>2$, and one formed at lower
redshift ($z_f<1$). The bands used trace, as mentioned, 
almost instantaneous star formation
more than the stellar mass growth studied in this paper. 
The lack of a detection of a secondary stellar mass growth episode at [3.6]
micron combined with its detection at shorter wavelengths implies
that the mass involved in such episodes is small. Quantification
of the mass involved will be reported elsewhere.

For the measurement of the stellar mass evolution, most of the
literature works compare mass estimates derived from different estimators at
different redshifts, because of the lack of similar measures (in the galaxy
rest-frame) over a large redshift range. Instead, our measures are
homogeneous and derived from a single estimator over the whole redshift range.
Our mass evolution estimate may rely on the
appropriateness of the adopted model (i.e. stellar mass growth history).  
However, we allow deviations from the model (several stellar mass growth
histories and non-null $Q$ values), but the data  rejected them. Furthermore,
literature approaches also rely on stellar mass growth models, and to a larger
degree than us, because they assume that the stellar mass growth history of {\it
each} individual galaxy is well described by a simple model, whereas we
assume that the above holds true for the {\it average} galaxy.  Averages,
by definition, are smoother and better described by a smooth model than
individual measures.

Our results on evolution
of the stellar mass in cluster galaxies
are in broad agreement with the literature. Bundy, Ellis and
Conselice (2005) find little evolution in the field from $z\sim1$ to
$z\sim0$, with particular emphasis on bright (massive) objects better
sampled in their (and in our) work. Our result also agrees with Kodama
\& Bower (2003) and Kodama et al. (2004), who claim that there has been
little evolution in $\mathcal{M^*}$, but with large uncertainties, by
comparing its value  measured in  cluster candidates  and the $z\sim0$
value (from Balogh et al. 2001), using heterogeneous mass estimators
and data.  Our sample of clusters at high redshift is larger than
theirs (we have 6 spectroscopically confirmed clusters at $z>0.99$, vs
3 candidate clusters in Kodama \& Bower 2003 and 5 
candidate clusters in Kodama et al. 2004), and our
comparisons use the same mass estimator and uniform data. 
Furthermore, spectroscopic observations  presented in a very recent
paper (Yamada et al. 2005) suggest that at least two of the Kodama et
al. (2003) clusters are instead line of sight superpositions.  The
cluster nature of all our clusters has been 
spectroscopically confirmed, and, for all
but three clusters, also confirmed through the detection
of the cluster x-ray emission.

\subsection{Early assembly and formation time of cluster galaxies}

The  evolution of the mass function of cluster galaxies only 
measures the  evolution of the galaxy population as a whole and does not
necessarily imply a direct correspondence to the evolution
of individual galaxies. For example, the constancy of the mass function
can be interpreted equally well as a combination of different and more
complicated evolution of individual galaxies, some of which grow
stellar mass (say, by mergers), and some that lose stellar mass.
However, such a possibility is unlikely, because it requires two
physical mechanisms with similar mass and time (i.e. redshift)
dependencies, otherwise the stellar mass function  would change. The
simplest interpretation, supported by the existence of very massive
($\mathcal{M}>10^{11.5} M_\odot$) galaxies in our clusters, is that the
mass assembly of most of galaxies in clusters (sampled with the
available data) was largely complete at $z>1.25$. 

Fundamental plane and colour studies (e.g. Bower et al. 1992;
Stanford et al. 1998; Kodama et al. 1998; Andreon 2003; Andreon et al. 2004a;
van Dokkum \& Stanford 2003 and references therein) suggest
that there is little recent star formation in early-type
or red galaxies but does not tell us whether these galaxies 
have been completely assembled. As long as early-type or
red galaxies are not a minority population in our clusters, 
the observed constancy of the mass function from $z=1.25$ to 
the present-day, as well as the old age of the stellar
populations, implies that these galaxies have
been completely assembled, and not only that their stars
are old.

\section{Summary}

We had measured the 3.6 $\mu$m luminosity evolution of about 1000
galaxies in 32 clusters at $0.2<z<1.25$, without any a priori
assumption about luminosity evolution, i.e. in a logically rigorous
way. Our data
match in number and in redshift distribution the cumulated 
literature effort thus far. The quality
of the data allows us to derive the LF and mass evolution 
homogeneously over the whole redshift range, using a
single estimator, at variance with previous determinations.
We found that the luminosity of our galaxies evolves as an old
and passively evolving population formed at high redshift without
any need for a further redshift-dependent evolution. Models with
a prolonged stellar mass growth
are rejected by the data with high confidence.  Data also reject
models in which the age of the stars is the same at all redshifts.   
Similarly, the characteristic mass function evolves as a
passively evolving stellar population formed at high redshift. The
Schechter function describes the galaxy luminosity function well.
The characteristic luminosity at $z=0.5$ is 16.30 mag with 
a 10 per cent uncertainty.

\begin{acknowledgements}
The author would like to warmly thank Giovanni Punzi 
for his statistical advice, and the anonymous referee for
his careful reading of the paper. This work used 
Minuit, by F. James. C. Lonsdale and S. Oliver are acknowledged
for useful discussions.
\end{acknowledgements}

\end{document}